\documentclass[12pt]{article}
\usepackage{amssymb,amsfonts}
\begin{document}
\hfil{{\bf Nonlinear Sigma Model, Zakharov-Shabat Method, and New Exact }}\hfil

\hfil{{\bf Forms of the Minimal Surfaces in ${\mathbb
R}^3$}}\hfil

\vskip0.5cm \hfil{{\bf E.Sh.Gutshabash}}\hfil
\vskip0.5cm \centerline{\small Faculty of Physics, St.Petersburg State University, Universitetskii pr.28, Petrodvorets }
\centerline{\small St.Petersburg, 198504, Russia} \centerline{\small e-mail:
gutshab@EG2097.spb.edu; e.gutshabash@spbu.ru} \vskip0.2cm  {\small General formulas for the construction of exact solutions of the equation of the minimal surface in
${\mathbb R}^3$, which appears in various physical problems, have been derived by the Zakharov-Shabat "dressing"\enskip method. Particular examples are considered.} \vskip0.7cm

{\bf 1.} The "area"\enskip functional plays an important role in many topical problems of theoretical and mathematical physics. In particular, according to current concepts [1], a free relativistic string is a continuum of points each moving along its world line in the D-dimensional spacetime. This motion occurs so that the area of the world sheet covered by the string in minimal (informal physical and geometric interpretation of this phenomenon was given in [2]).

We begin with the area functional in the form

$$
S[f]=\int d^{D-1}y \sqrt {{|\mathrm det\:g^{(0)}}|},           \eqno(1)
$$
where $g^{(0)}=||g^{(0)}_{ij}(y)||, \:\:g^{(0)}_{ij}(y)$ is induced Riemann metric on a hypersurface,
$y=(y^{1},\:y^{2},...,y^{D-1}),\: d^{D-1}y=dy^1\bigwedge
dy^2...\bigwedge dy^{D-1},\:(y,\:y^{D})\in \Omega \subset {\mathbb
R}^{D}$, and $y^D=f(x)$ is the smooth hypersurface (it is assumed that the function $f(y)$ is real-valued).

Then, the Euler-Lagrange equation (critical point) for Eq.(1) is written as

$$
{\mathrm div}\:(\frac
{f_{y^{i}}}{\sqrt{1+\sum_{l=1}^{D-1}(f_{y^{l}})^2}})=0.    \eqno(2)
$$
This equation is hardly integrable for arbitrary D value. As will by shown below, this equation  with $D=3$ has a Lax representation and, therefore, belongs to the class of completely integrable systems. The aim of this work is to construct its new exact solutions.

Equation (2) for $D=3$ can be represented in the following more distinct form by setting $y^{1}=t,\:y^{2}=x$:

$$
(1+f_t^2)f_{xx}-2f_{xt}f_xf_t+(1+f_x^2)f_{tt}=0. \eqno(3)
$$
This two-dimensional nonlinear elliptic equation has long been known in differential geometry. It coincides with the condition of zero average curvature of the surface and is referred to as  the "equation of the minimal surface"\enskip  (MS) (in
${\mathbb R}^3$) {\footnote{If the contour $\partial\Omega$, on which the surface is spanned is closed, it is reasonable to consider a Dirichlet problem for this equation. This boundary value problem is called in mathematics Plateau's problem \enskip [4]. Further details concerning Eq.(3) can be found, e.g. in [5] and a number of its different representations, as well as a series of local conservation laws, were given in [6].}}. Equation (3)
appears, e.g., at some reductions of the ${\bf
n}$-field model [7], in the stationary and dispersionless limit of the system of coupled nonlinear Shr\"{o}dinger equation [8], and some other physical problems. In this sense, it has a certain universal significance. It is worth emphasizing that the so-called integral Enneper-Weierstrass representations [9], as well as linearization of Eq.(3) by means of the Legendre or hodograph transforms [10], generally give only some parametrizations of the solution or their limited set (the plane in ${\mathbb R}^3$, helicoid, catenoid, Scherk surface, and Enneper surface [11] are among the known solutions of Eq. (3)). For this reason, the approach developed below on the basis of direct integration of Eq. (3) seems to be only possible method for solving the formulated problem.

It is also noteworthy that it is geometrically obvious that the image of the minimal surface is invariant under rotation, shift and similarity transformation.

{\bf 2.} Equation (3) can be represented in the form of the compatibility condition for Lax pair [12]

$$
\Psi_x=\frac{1}{\lambda^2+1}[\lambda(g^{11}I_1+g^{12}I_2)-I_2]\Psi \equiv U\Psi,    \eqno(4a)
$$
$$
\Psi_t=-\frac{1}{\lambda^2+1}[\lambda(g^{21}I_1+g^{22}I_2)+I_1]\Psi \equiv V\Psi,    \eqno(4b)
$$
where

$$
g=\left(\matrix
{u_{tt} & u_{tx} \cr u_{tx} & u_{xx}}\right),\:\:{\mathrm det}\:g=1,\:\:g=g^T,\:\:g^2 \ne I,\:g^{-1}=\{g^{kl}\},
\eqno(5)
$$

$$
u_{xx}=\frac{1+f_x^2}{\cal L},\:\:u_{tt}=\frac{1+f_t^2}{\cal L},\:\:u_{xt}=\frac{f_x
f_t}{\cal L},\:\:{\cal L}=\sqrt{1+f_x^2+f_t^2},  \eqno(6)
$$
$I_1=g^{-1}g_t,\:I_2=g^{-1}g_x,\:\Psi=\Psi(x, t, \lambda) \in {\mathrm Mat}(2, {\mathbb C}), \:\lambda \in {\mathbb C}$ is the spectral parameter, and ${\cal L}$ is the Lagrangian density of Eq. (3). The unimodularity condition for the matrix $g$ means that the function $u=u(x, t)$ satisfies a Monge-Ampere elliptic equation

$$
u_{tt}u_{xx}-u_{tx}^2=1,  \eqno(7)
$$
so that ${\mathrm Tr} I_1={\mathrm Tr} I_2=0$ and $I_1,\:I_2 \in sl(2)$, where $sl(2)$ is the Lie algebra of the $SL(2)$ group. Then, consistency requirement  (4), together with the identity

$$
I_{2t}-I_{1x}-[I_2, I_1]=0    \eqno(8)
$$
leads to a nonlinear sigma-model of the form $(\alpha,\:\beta=1,2)$

$$
\partial_{\alpha}(g^{\alpha \beta}g^{-1}\partial_{\beta}g)=0,   \eqno(9)
$$
which is equivalent to Eq.  (3).

Exact solutions of Eq. (3) for the discrete spectrum of associated linear system (4a) are constructed with the use of the Zakharov-Shabat dressing procedure [13] (in the variant used in [14]-[15] and also in [16]){\footnote {The boundary value problem for nonlinear elliptic equations on a half-plane, the entire plane, and a quarter plane was solved by means of the inverse scattering problem method in [16]-[17], [18], and [19], respectively.}}.

First, it directly follows from Eq. (4a) that

$$
g=\Psi^{-1}(0).     \eqno(10)
$$
Then, we know a certain bare (background) solution $f_0=f_0(x,t)$ of Eq. (3) and this solution, according to Eq. (4),
corresponds to the matrix solution $\Psi_0=\Psi_0(x, t, \lambda)$, and  $g=g_0(x, t)$. We set
$\Psi=\chi \Psi_0$, where $\chi=\chi(x, t, \lambda)\in \mathrm
{Mat(2, {\mathbb C})}$, assuming the canonical normalization $\chi(\infty)=I$ (where $I$ is $2\times 2$ identity matrix). From the properties of the matrix $U$ ($\sigma_i,\:i=1, 3$ are the standard Pauli matrix and the asterisk stands for Hermitian conjugation),

$$
U(\lambda)={\bar U}(\bar {\lambda}),\:\:U(\lambda)=-\sigma_2U^{T}(\lambda)\sigma_2,\:\:gU(\lambda)=U^{*}({\bar\lambda})g,
 \eqno(11)
$$
the involutions (symmetries) for the function $\Psi$ and $\chi$ follows in the form

$$
{\bar \Psi}({\bar \lambda})=\Psi(\lambda)D_1(\lambda),\:\:\Psi^{T}(\lambda)\sigma_2\Psi(\lambda)=
D_2(\lambda),\:\:g\Psi(\lambda)D_3(\lambda)=\Psi^{-1T}(-\frac{1}{\lambda}),   \eqno(12)
$$
and
$$
{\bar \chi}(\bar {\lambda})=\chi(\lambda),\:\:\chi^{T}(\lambda)=\sigma_2\chi^{-1}(\lambda)\sigma_2,\:\:
\chi^{-1}(0)\chi(\lambda)\Psi_0(0)=\Psi_0(0)\sigma_2\chi(-\frac{1}{\lambda})\sigma_2,            \eqno(13)
$$
where $D_i(\lambda)$ are arbitrary matrices ($i=1,3$) such that  $D_1(\lambda){\bar D}_1({\bar \lambda})=I,\:D_2^{T}(\lambda)=-D_2(\lambda)$.
Taking into account Eqs. (13), we seek $\chi$ and $\chi^{-1}$ in the form

$$
\chi=I+\frac{P_1}{\lambda-\lambda_1}+\frac{{\bar P}_1}{\lambda-{\bar \lambda}_1}+
\frac{R_1}{\lambda+\lambda_1^{-1}}+\frac{{\bar R}_1}{\lambda+{\bar \lambda}_1^{-1}},
$$
$$
\eqno(14)
$$
$$
\chi^{-1}=I+\frac{\sigma_2P_1^T\sigma_2}{\lambda-\lambda_1}+\frac{\sigma_2
P_1^{*}\sigma_2}{\lambda-{\bar \lambda}_1}+
\frac{\sigma_2R_1^{T}\sigma_2}{\lambda+\frac{1}{\lambda_1}}+\frac{\sigma_2R_1^{*}\sigma_2}{\lambda+{\bar \lambda}_1^{-1}},
$$
where $P_1=P_1(x, t),\:R_1=R_1(x, t) \in {\mathrm Mat}(2, {\mathbb C})$ are as yet unknown function and $\lambda_1 \in {\mathbb C}$ is the simple pole of the function $\chi$ such that $\lambda_1=\lambda_{1R}+\lambda_{1I},\:\lambda_{1R}\ne 0$ and
$\lambda_{1I}\ne 0$.  In terms of $P_1=a><p\sigma_2$ and
$R_1=b><q\sigma_2$, the condition  $\chi \chi^{-1}=I$
after equating combination of residues at the same poles to zero gives the following system of linear equations for the vectors $a>,\:{\bar a}>,\:b>$  and ${\bar b}>$:

$$
\frac{1}{\lambda_1-{\bar \lambda}_1}{\bar a}><{\bar p}\sigma_2p>-\frac{1}{\lambda_1+\lambda_1^{-1}}b><q\sigma_2p>+\frac{1}{\lambda_1+{\bar \lambda_1}^{-1}}
{\bar b}><{\bar q}\sigma_2p>=p>,
$$

$$
-\frac{1}{\lambda_1-{\bar \lambda}_1}a><p\sigma_2{\bar p}>+\frac{1}{{\bar \lambda}_1+\lambda_1^{-1}}b><q\sigma_2{\bar p}>-\frac{1}{\bar \lambda_1+\bar \lambda_1^{-1}}
{\bar b}><{\bar q}\sigma_2{\bar p}>=-{\bar p}>,
$$
$$
\eqno(15)
$$
$$
\frac{1}{\lambda_1+\lambda_1^{-1}}a><p\sigma_2q>-\frac{1}{{\bar \lambda}_1+\lambda_1^{-1}}{\bar a}><{\bar p}\sigma_2q>+\frac {1}{\bar \lambda_1^{-1}-\lambda_1^{-1}}
{\bar b}><{\bar q}\sigma_2q>=q>,
$$
$$
-\frac{1}{\lambda_1+{\bar \lambda}_1^{-1}}a><p\sigma_2{\bar q}>+\frac{1}{{\bar \lambda}_1+\bar \lambda_1^{-1}}{\bar a}><{\bar p}\sigma_2{\bar q}>-\frac{1}{\bar \lambda_1^{-1}-\lambda_1^{-1}}b><q\sigma_2{\bar q}>=-{\bar q}>.
$$
The requirement of the absence of doubles poles is reduced to the conditions

$$
<p\sigma_2p>=<q\sigma_2q>=0.         \eqno(16)
$$
In addition, it is easy to verify that

$$
<p\sigma_2{\bar p}>=-<{\bar p}\sigma_2p>=0,\:\:\:<q\sigma_2{\bar q}>=-<{\bar q}\sigma_2q>=0,
$$
$$
\eqno(17)
$$
$$
<p\sigma_2q>={\overline {<q\sigma_2p>}},\:\:\:<p\sigma_2{\bar q}>=-{\overline {<{\bar p}\sigma_2q>}}.
$$
Solving system (15) with allowance for Eq. (16) and (17), we obtain the matrices $P_1$ and $R_1$ in the form

$$
P_1=\frac{1}{|d_1|^2|l_1|^2-|d_2|^2|l_2|^2}[{\bar d_1}l_1q> <p-{\bar d_2}l_2{\bar q}> <p]\sigma_2,
$$
$$
\eqno(18)
$$
$$
R_1=-\frac{1}{|d_1|^2|l_1|^2-|d_2|^2|l_2|^2}[{\bar d_1}{\bar l_1}p> <q+d_2l_2{\bar p}> <q ]\sigma_2,
$$
where

$$
l_1=<q\sigma_2p>,\:\:l_2=<{\bar q}\sigma_2p>,\:\:d_1=\frac{1}{\lambda_1+\lambda_1^{-1}},\:\:d_2=\frac{1}{\lambda_1+{\bar \lambda_1}^{-1}}.
$$
In this case, $p> <q=(q> <p)^{*},\:{\bar q}> <p =({\bar p}> < q)^{T}$, and, in view of Eqs. (16) and (17), $P_1|p>=0$ and $R_1|q>=0$, so, that
$|p> \in {\mathbb Ker P}_1,\:|q> \in {\mathbb Ker R}_1$.

The next step in the dressing method is the determination  of the dependence of the vectors $|p>$ and $|q>$ on the variables  $x$ and $t$.
To this end, system (4) is rewritten in terms of the solutions $\chi$ and $\chi^{-1}$; since the functions  $U(\lambda)$ and $V(\lambda)$  have poles only at the points $\pm i$, we impose the conditions ($U_0=\Psi_{0x}\Psi_0^{-1},\:
V_0=\Psi_{0t}\Psi_0^{-1}$):

$$
{\mathrm Res}_{|\lambda=\lambda_1}\{\chi(-\partial_x+U_0(\lambda))\chi^{-1}\}=0,\:\:\:
{\mathrm Res}_{|\lambda=\lambda_1}\{\chi(-\partial_t+V_0(\lambda))\chi^{-1}\}=0.  \eqno(19)
$$
According to Eq. (14), near the point $\lambda_1$

$$
\chi(\lambda)=\frac{P_1}{\lambda-\lambda_1}+A_0(\lambda),\:\:\: \chi^{-1}(\lambda)=\frac{\sigma_2 P_1^{T}\sigma_2}{\lambda-\lambda_1}+B_0(\lambda),   \eqno(20)
$$
where $A_0(\lambda)$ and $B_0(\lambda)$ are the functions analytic at the point $\lambda_1$, such that $B_0(\lambda)=\sigma_2A_0^T(\lambda)\sigma_2$.
The substitution of the Eq. (20) into Eq. (19) gives

$$
A_0(\lambda_1)(\partial_x-U_0)B_0=0,     \eqno(21)
$$
and
$$
A_0(\lambda_1)(\partial_x-U_0)\sigma_2P_1^T\sigma_2+P_1(\partial_x-U_0)\sigma_2A_0^T(\lambda_1)\sigma_2=0,
$$
$$
\eqno(22)
$$
$$
A_0(\lambda_1)(\partial_t-V_0)\sigma_2P_1^T\sigma_2+P_1(\partial_t-V_0)\sigma_2A_0^T(\lambda_1)\sigma_2=0.
$$
These equalities are obviously satisfied if $(\partial_x-U_0)\sigma_2P_1^T=(\partial_t-V_0)\sigma_2P_1^T=0$.
Comparing them with Eqs. (4a) and (4b), we obtain ($|\alpha_1>$ is a constant complex vector)

$$
p>=\Psi_0(x, t, {\bar\lambda}_1)|\alpha_1>.      \eqno(23)
$$
The same result can be obtained by repeating the above calculations near the point ${\bar \lambda_1}$.
The dependence of the vector $q>$ is obtained similarly by applying the same reasons near the point $-1/\lambda_1$ or $-1/{\bar
\lambda}_1$. Then ($|\alpha_2>$ is a constant complex vector),

$$
q>=\Psi_0(x, t, -\frac{1}{{\bar \lambda}_1})|\alpha_2>.  \eqno(24)
$$

To close the above procedure, it is necessary to obtain a formula for the reconstruction of the "potential" \enskip $f$, and to construct the solution $\Psi_0(x, t, \lambda)$. To this end, it is easy to derive the following equation for the Lagrangian density of the "dressed"\enskip solution from Bianchi relations (6) and formula (10):

$$
{\cal L}^2-({\mathrm Tr} \Psi(0)){\cal L}+1=0.  \eqno(25)
$$
Therefore (under the assumption that $|{\mathrm Tr} \Psi(0)|\geq 2$),

$$
|\nabla f|^2=\frac{1}{2}{\mathrm Tr} \Psi(0)\Bigl [{\mathrm Tr} \Psi(0)\pm \sqrt{{\mathrm Tr}^2\Psi(0)-4}\:\Bigr ]-2.         \eqno(26)
$$
The scalar functions $Q_1=Q_1(x, t, \lambda_1),\:Q_2=Q_2(x, t, \lambda_1)$ are introduced as
$$
Q_1=\Psi_{11}(0)=\frac{(1+f_{0x}^2)\chi_{11}(0)-f_{0x}f_{0t}\chi_{12}(0)}{{\cal L}_0}=\frac{E^{(0)}\chi_{11}(0)-F^{(0)}\chi_{12}(0)}{\sqrt{E^{(0)} G^{(0)}-F^{(0)2}}},
$$
$$
\eqno(27)
$$
$$
Q_2=\Psi_{22}(0)=\frac{(1+f_{0t}^2)\chi_{22}(0)-f_{0x}f_{0t}\chi_{21}(0)}{{\cal L}_0}=\frac{G^{(0)}\chi_{22}(0)-F^{(0)}\chi_{21}(0)}{\sqrt{E^{(0)} G^{(0)}-F^{(0)2}
}},
$$
 where $\Psi(0)=\{\Psi_{ij}(0)\},\:\chi(0)=\{\chi_{ij}(0)\},\:{\cal L}_0^2=1+f_{0x}^2+f_{0t}^2=E^{(0)}G^{(0)}-F^{(0)2}$, and $E^{(0)},\:F^{(0)}$, and $G^{(0)}$ are the coefficients of the first quadratic form of the "initial"\enskip ("bare") surface. If
 $\Psi_0(0)$ is real-valued,  $Q_1$ and $Q_2$ are also real-valued, because $\chi(0)={\bar \chi(0)}$ according to Eq. (14). Taking into account Eqs. (10) and (26), we obtain

$$
f_x^2=\frac{Q_1}{2}[Q_1+Q_2\pm \sqrt{(Q_1+Q_2)^2-4}]-1 \equiv 2Q_1e^{Q_3}-1 \equiv \Delta_1^2,
$$
$$
\eqno(28)
$$
$$
f_t^2=\frac{Q_2}{2}[Q_1+Q_2\pm \sqrt{(Q_1+Q_2)^2-4}]-1 \equiv 2Q_2e^{Q_3}-1 \equiv \Delta_2^2,
$$
where $Q_1+Q_2=4\cosh Q_3$. Direct calculation shows that $\Delta_{1t}=\Delta_{2x}$. This means that the differential 1-form
$\omega=\Delta_1 dx+\Delta_2 dt$ is exact. Consequently, the curvilinear integral

$$
f(x, t)=\pm \int_{(x_0, t_0)}^{(x, t)} \Delta_1(x^{\prime}, t)\:dx^{\prime}+
\Delta_2(x, t^{\prime}) \:dt^{\prime}=\pm \int_{(x_0, t_0)}^{(x, t)} \sqrt{2Q_1e^{Q_3}-1}dx+\sqrt{2Q_2e^{Q_3}-1}dt,  \eqno(29)
$$
where $(x_0, t_0)\in \Omega$ is a certain point, is independent of integration path and the presence of both signs corresponds to the invariance of Eq. (3) under the change $f \to -f$.
Expressions (27) and (28), together with  Eq. (29), specify a "single-soliton"\enskip  solution of Eq.3 against the background of the solution  $f_0(x,t)$. In this case, from the geometric point of view, the dressing procedure is reduced to a nonlinear transformation of the coefficients of the first quadratic form of the initial surface.

To determine $\Psi_0(x,t, \lambda)$ we note that the system (4) after some algebra can be rewritten in the form

$$
\Psi_{0x}=\frac{1}{\lambda^2+1}[-\lambda G_{0x}-G_0^{-1}G_{0x}]\Psi_0,\:\:
\Psi_{0t}=\frac{1}{\lambda^2+1}[-\lambda G_{0t}-G_0^{-1}G_{0t}]\Psi_0 ,    \eqno(30)
$$
where $G_0=i\sigma_2g_0$, so that $G_0^2=-I$ (in view of Eq. (7)). This system has the solution

$$
\Psi_0(x, t, \lambda)=ic_0(\lambda)[G_0^{-1}+(\lambda-\lambda^3)I+\lambda^2G_0]\sigma_2,          \eqno(31)
$$
where $c_0(\lambda)$ is the factor, determined by the requirement ${\mathrm det}\:\Psi_0(\lambda)=1$. Then, in view of the identity $\sigma_2g_0\sigma_2=g_0^{-1}$, Eq. (31) can be represented in terms of the derivations of the function $u_0$ (initial solution of Eq. (7)):

$$
\Psi_0(\lambda)=\{\Psi_{0ij}(\lambda)\}=\frac{1}{\sqrt{1+\lambda^2}}\left(\matrix
{u_{0xx} & \lambda-u_{0tx} \cr -(\lambda+u_{0tx}) & u_{0tt}}\right), \:\:\Psi_0(0)={\bar \Psi_0}(0).  \eqno(32)
$$

{\bf 3.} We define the vectors $|\alpha_i>,\:i=1,2$, entering into Eqs. (23) and (24) as $|\alpha_i>=(\mu_i, \nu_i)^{T}$,
where $\mu_i,\:\nu_i \in {\mathbb C}$ are arbitrary numbers. In order to simplify the formulas, it is convenient to set $|\alpha_1>=\mu_{10}(1+i\mu_0)(1, -i)^T$ and $|\alpha_2>=\mu_{20}(1+i\mu_0)(1,i)^T$, where $\mu_{i0},\:\mu_0  \in {\mathbb R}$ are the parameters. Furthermore, the parameter $\lambda_1$ is taken on a unit circle, i.e., $\lambda_1=e^{i\theta_1},\:\theta_1\neq 0,\:\pi/2,\:\pi,\:3\pi/2,\:2\pi$, which corresponds to "kink"\enskip excitations in spectral problem (4a) [17]. Then, in the general situation, i.e., for any MS  $f_0=f_0(x,t)$, according to Eqs. (23), (24), and (32) we obtain ($d_1=1/(2\cos \theta_1),\:d_2=(1/2)e^{-i\theta_1}$)

$$
|p>=(p_1, p_2)^T=\frac{\mu_{10}(1+i\mu_0)e^{\frac{i\theta_1}{2}}}{{\cal L}_0\sqrt{2|\cos \theta_1|}}\left(\matrix
{{\cal L}_0({\cal L}_0-ie^{-i\theta_1})+f_{0t}(f_{0x}-1) \cr -e^{-i\theta_1}{\cal L}_0-f_{0x}f_{0t}(1+i)}\right),
$$
$$
|q>=(q_1, q_2)^T=\frac{{\bar \lambda_1}\mu_{10}(1+i\mu_0)e^{\frac{i\theta_1}{2}}}{{\cal L}_0\sqrt{2|\cos \theta_1|}}\left(\matrix
{{\cal L}_0[1+e^{-2i\theta_1}(1-i)]+i(e^{-i\theta_1}f_{0x}f_{0t}-1-f_{0x}^2) \cr e^{-i\theta_1}[i{\cal L}_0(2\cos \theta_1-e^{-i\theta_1})-f_{0x}f_{0t}]-1-f_{0t}^2}\right),
$$
$$
l_1=-l_{10}\frac{1+{\cal L}_0^2-2{\cal L}_0\sin \theta_1}{{\cal L}_0|\cos \theta_1|},\:\:l_2= l_{10}\frac{f_{0x}^2-f_{0t}^2+2if_{0x}f_{0t}}{{\cal L}_0},\:\:l_{10}=\mu_{10}\mu_{20}(1+\mu_0^2),
$$
$$
\eqno(33)
$$
$$
\Delta=\Delta({\cal L}_0,\:\theta_1)=|d_1|^2|l_1|^2-|d_2|^2|l_2|^2=\frac{l_{10}^2}{4{\cal L}_0^2}\frac{(1+{\cal L}_0^2-2{\cal L}_0\sin \theta_1)^2-
(1-{\cal L}_0^2)^2\cos^4 \theta_1}{\cos^4 \theta_1} \neq 0,
$$
$$
A_1=q><p=({\bar p}_1 |q>,\:{\bar p}_2 |q>),\:
A_2={\bar q}><p=({\bar p}_1 |{\bar q}>, {\bar p}_2 |{\bar q}>),\:A_1,\:A_2 \in {\mathrm Mat(2, {\mathbb C})},
$$
$$
\chi(0)=I-\frac{2}{\Delta}{\mathrm Re} \{
[e^{-i\theta_1}({\bar d}_1l_1A_1-{\bar d}_2l_2A_2)-e^{i\theta_1}({\bar d}_1{\bar l}_1A_1^{*}+ d_2l_2A_2^{T})]\sigma_2 \}.
$$

As a result, solution (29) is very lengthy and, for this reason, is not given herein a more explicit form. It is nonsingular and depends on four real parameters $\mu_0,\:\mu_{10},\mu_{20},\:\theta_1$.

Two particular examples of implementations of the above relations are as follows.

i). Let $f_0(x, t)=0$; i.e., the procedure of dressing of the trivial solution ("horizontal"
\enskip plane) is considered. In this case,  ${\cal L}={\cal L}_0=1,\:
u_{0xx}=u_{0tt}=1,\:u_{tx}=0,\:g_0=I,\:I_1=I_2=0$, and $\Psi_0(\lambda_1)=(1/\sqrt{2|\cos \theta_1|})e^{-i\theta_1/2}[I+i(\cos \theta_1\sigma_2+\sin \theta_1\sigma_1)]$. The matrix $\chi(0)$ and, therefore, the functions $Q_1$ and $Q_2$ are constants. According to Eq.  (29), this means that the solution $f(x, t)$ is a linear function of $x$ ш $t$, i.e., an "inclined" \enskip plane, which can be transformed to the initial position by rotation and shift transformations. This result corresponds to the classical Bernshtein result [20]: the image of the minimal surface over the entire plane is the plane itself. Thus, we have a certain "exclusion principe"\enskip for new solutions over the entire plane.

The situation for the bare solution in the form of an inclined plane is similar, i.e., $f_0(x,t)=a_0x+b_0t+c_0$, where $a_0,\:b_0,\:c_0 \in {\mathbb R}$ are constants.

ii). Let $f_0(x,t)$ be the MS different from a plane, e.g., a helicoid $f_0=x\tan t,\:x \in {\mathbb R},\:t \in {\mathbb R}\setminus(\pi/2)(2k+1),\:k=0,\pm 1,\pm 2, \ldots $. In this situation, $ f_{0x}=\tan t,\:f_{0t}=x/\cos^2 t,\:{\cal L}_0=(\sqrt{x^2+\cos^2 t})/\cos^2 t$, and evolution in $x$ and $t$ is nontrivial (at least for $f_x^2$ and $f_t^2$) and is determined according to the first  two formulas in Eqs. (33). In this case,

$$
l_1=-l_{10}\frac{x^2+\cos^2 t(1+\cos^2 t)-2\sin \theta_1\cos^2 t\sqrt{x^2+\cos^2 t}}{|\cos \theta_1|\sqrt{x^2+\cos^2t}},
$$
$$
l_2=l_{10}\frac{\sin^2 t -x^2+2ix \tan t}{\sqrt{\cos^2 t+x^2}},                    \eqno(34)
$$
$$
\Delta=\frac{l_{10}^2}{\cos^6 t\cos^4 \theta_1}
\frac{(\cos^4 t+\cos^2 t+x^2-2\sin \theta_1\sqrt{\cos^2 t+x^2})^2-(\cos^4 t-\cos^2 t-x^2)^2\cos^4 \theta_1}{\cos^2 t+x^2}.
$$
\vskip0.5cm

To conclude, it is noteworthy that the procedure used above makes it possible to simultaneously solve three equations: equation of MS (3), sigma-model (9), and Monge-Ampere equation (7).

I am grateful to A.R. Its and P.P. Kulish for stimulated discussions and support.

\vskip1.5cm

\hfil {\bf {REFERENCES}}\hfil

\vskip1.5cm

\vskip0.3cm  1. \parbox [t] {12.7cm}
       {L.Brink and M.Henneaux, -{\em  Principles of String Theory,} Plenum Press, New York, 1988; Mir, Moscow, 1991 (in Russian).}

\vskip0.3cm  2. \parbox [t] {12.7cm}
       {L.D. Nikitina, I.N.Nikitin, and S.V.Klimenko, Elektr.Zh. "Issledovano v Rossii", 6,  404-417, (2003).}

\vskip0.3cm  3. \parbox [t] {12.7cm}
       {B.A. Dubrovin, S.P. Novikov, and A.T. Fomenko, - {\em Modern Geometry,}
       Nauka, Moscow, 1999 (in Russian).}

\vskip0.3cm  4. \parbox [t] {12.7cm}
       { A.A. Tuzhilin and A.T. Fomenko, - {\em Elements of the Geometry and Topology of Minimal Surfaces in Three-Dimensional Space}, Nauka, Moscow, 1991 (in Russian); Amer.Math.Soc., New York, 2005.}

\vskip0.3cm  5. \parbox [t] {12.7cm}
       {R.Osserman, (Ed.), - {\em Geometry V,  Minimal surfaces,}  Springer, Berlin, 1997.}

\vskip0.3cm 6. \parbox [t] {12.7cm}
       {E.Sh.Gutshabash, - Zapiski Nauchnych Seminarov POMI,  {\bf 374},  121-135, (2010).}

\vskip0.3cm   7. \parbox [t] {12.7cm}
       {A.B. Borisov, - Dokl.Akad.Nauk, {\bf 389}, 1, (2003).}

 \vskip0.3cm   8. \parbox [t] {12.7cm}
       {A.Moro, - arXiv: 0808.1235. (2008).}

 \vskip0.3cm  9. \parbox [t] {12.7cm}
       {V.Bliashke, - {\em Introduction to Differential Geometry, Regulyar. Khaotich. Dinamika, Izhevsk,}  2007 (in Russian).}

\vskip0.3cm  10. \parbox [t] {12.7cm}
       {A.D.Polyanin, V.F.Zaitsev, and A.I.Zhurov, - {\em Solution Methods for Nonlinear Equations of Mathematical Physics and Mechanics,}  Fizmatlit, Moscow,  2005 (in Russian).}

\vskip0.3cm  11. \parbox [t] {12.7cm}
       {U.Dierkes, S.Hildebrant, F.Sauvigny, -  {\em Minimal surfaces. A Series of Comprehensive Studies in Mathematics,} vol. 239, Springer, Berlin, 2010.}

\vskip0.3cm  12. \parbox [t] {12.7cm}
       {J.C.Brunelli, M.Gurses, K.Zheltukhin, - Rev. Math. Phys, {\bf 13}, 529, (2001).}

\vskip0.3cm  13. \parbox [t] {12.7cm}
       {V.E.Zakharov, A.B. Shabat, - Funkts. Analiz Prilozh.,  {\bf 8}, 43, (1974).}

 \vskip0.3cm 14. \parbox [t] {12.7cm}
       {A.V.Mikhailov, A.I.Jaremchuk, - Nucl.Phys.B, {\bf 202},
       508-522, (1982).}

\vskip0.3cm  15. \parbox [t] {12.7cm}
       {F.Combes, H.J. de Vega, └.V.Mikhailov, and N.Sanchez, - Phys. Rev. D, {\bf 50},
4, 2754, (1993).;  arXiv:hep-th/9310073.}

\vskip0.3cm  16. \parbox [t] {12.7cm}
       {E.Sh.Gutshabash, V.D.Lipovskii, and S.S. Nikulichev, - Teor. Mat. Phys., {\bf 115}, 323, (1998).}

\vskip0.3cm  17. \parbox [t] {12.7cm}
       {E.Sh. Gutshabash and V.D. Lipovskii, - Zapiski Nauchnych Seminarov POMI {\bf 180}. 53-62, (1990).}

\vskip0.3cm  18. \parbox [t] {12.7cm}
{A. B. Borisov, V. V. Kiseliev, - Inverse Problems {\bf 5}, 959, (1998).}

\vskip0.3cm  19. \parbox [t] {12.7cm}
{B. Pelloni, - Theor. Math. Phys,  {\bf 160}, 1031, {2009}.}

\vskip0.3cm  20. \parbox [t] {12.7cm}
{S.N.Bernshtein,- {\em Collection of Scientific Works (Akad.Nauk SSSR),}  vol.3, ╠oscow, 1960 (in Russian).}

\end{document}